\documentclass[preprint,pra,amsmath,amssymb,aps,showpacs]{revtex4}
\usepackage[utf8]{inputenc}
\usepackage{graphicx}
\usepackage{bm}
\usepackage{subfigure}
\usepackage{amsmath}

\newcommand{\epl}{Europhys. Lett.\ }

\newcommand{\pr}{Phys. Rev.\ }

\newcommand{\jpa}{J. Phys. A\ }

\newcommand{\njp}{New J. Phys.\ }
\newcommand{\etal}{{\em et al. }}

\newcommand{\tr}{\mbox{tr}}

\newcommand{\UQ}{School of Mathematics and Physics, University of Queensland, Brisbane, 
QLD 4072, Australia.}

\newcommand{\CAEN}{Ecole Nationale Sup\'erieur d’Ing\'enieur de Caen, Caen 14000, France}

\begin{document}

\title{Finite size effects and equilibration in Bose-Hubbard chains with central well dephasing}

\author{F. Martinet}
\affiliation{\CAEN}

\author{M.~K. Olsen}
\affiliation{\UQ}

\date{\today}

\begin{abstract}

We investigate Bose-Hubbard chains in a central depleted well configuration, with dephasing in the middle well. We look at equilibration of populations, pseudo-entropy, and entanglement measures. Using stochastic integration in the truncated Wigner representation, we find that the initial quantum states of the occupied wells has an influence on the subsequent dynamics, and that with more than three wells, the chains do not reach a full equilibrium, with edge effects becoming important, and the time to reach the steady state becoming longer. The evolutions with and without phase diffusion are qualitatively different. We find no convincing evidence of entanglement in the final states with phase diffusion. Although at least one accepted measure indicates the presence of mode entanglement, we are easily able to show that it can give ambiguous predictions.

\end{abstract}


\maketitle

\section{Introduction}
\label{sec:intro}

There have recently been many experimental advances with regard to the manipulation of ultra-cold bosons in optical traps.  Examples include  ``potential painting'', which allows for the dynamical fabrication of almost arbitrary potentials~\cite{painting,tylerpaint}, and the use of either an electron beam~\cite{NDC} or optical methods~\cite{Weitenberg} to empty chosen wells of an optical lattice system. Combined with the well-known techniques of scattering length change via Feshbach resonance~\cite{Feshbach}, such methods allow fine-tuned, dynamical control over nearly all the experimental parameters of an optical lattice system.The experimental example that inspires this work is that of Labouvie~\etal~\cite{NDC}, who used an electron beam to remove a proportion of the atoms from a given well and measured the filling rate from the neighbouring wells as a function of chemical potential difference. 

Here, we examine the dynamics of Bose-Hubbard~\cite{BHmodel,Jaksch} chains in which the middle well is initially empty, so that the initial difference in chemical potential is fixed. A similar system with three wells, but without collisional dephasing has previously been analysed by Penna~\cite{CDW}, in what he named the central depleted well regime. Aspects of a similar system have also been analysed by Kordas~\etal\cite{Kordas}, who found that phase diffusion could enhance tunneling and damp coherent oscillations in the tunneling. Kordas and others have also analysed the influence of atomic losses on a Bose-Hubbard system, outlining various useful theoretical approaches~\cite{Kordas2}. The existence of negative differential conductivity and the influence of initial quantum states in a three-well chain has also been theoretically analysed by Olsen and Corney~\cite{ourNDC}.

We proceed via the numerical simulation of the quantum dynamics of Bose-Hubbard chains using stochastic equations derived in the truncated Wigner representation~\cite{Robert,Steeletal}. Our  simulations quantify the contributions of dephasing, chain length, and initial quantum states to the dynamics of the systems. We find that the inclusion of dephasing in the middle well  qualitatively alters the steady-state of the chains, in terms of both the number distributions and the quantum statistics. We also find that the initial quantum states of the occupied wells can have a large effect on both the dynamics and the final states, demonstrating that the preparation of an experimental system can have a drastic effect on the subsequent dynamics.

\section{Physical model and equations of motion}
\label{sec:model}

The systems we investigate are relatively simple, being Bose-Hubbard chains in a linear configuration which is symmetric about an initially empty middle well. Each of the wells can contain a single atomic mode, which we will treat as being in the
lowest energy level, and atoms can tunnel into the nearest-neighbour potential. As in the Labouvie paper, we include a phenomenological dephasing term in the middle well~\cite{NDC,ourNDC}, which mimics the effects of scattering between different radial modes in the initially depleted well.

Following the approach taken by Milburn \etal~\cite{BHJoel}, while
generalizing this to three and more wells~\cite{Nemoto,BH42011,BHrelax,BHBS,BHspread}, we can write the Hamiltonian for an $n$-well chain as
\begin{equation}
{\cal H}_{n} = \hbar\chi\sum_{j=1}^{n}\hat{a}_{j}^{\dag\;2}\hat{a}_{j}^{2}+\hbar J\sum_{j=1}^{n-1}\left(\hat{a}_{j}\hat{a}_{j+1}^{\dag}+\hat{a}_{j}^{\dag}\hat{a}_{j+1} \right),
\label{eq:Ham}
\end{equation}
where $\hat{a}_{j}$ and $\hat{a}_{j}^{\dag}$ are the bosonic annihilation and creation operators for atoms in the mode of the $n$th well, $J$ is the tunnel coupling between the wells, and $\chi$ is the collisional nonlinearity due to $s$-wave scattering. The dephasing in the middle well is represented by the Liouvillian superoperator acting on the reduced density matrix for the central well,
\begin{equation} 
{\cal L}\rho_{m} = \frac{\Gamma}{2}\left(2\hat{a}_{m}^{\dag}\hat{a}_{m}\rho_{m}\hat{a}_{m}^{\dag}\hat{a}_{m}  - \hat{a}_{m}^{\dag}\hat{a}_{m}\hat{a}_{m}^{\dag}\hat{a}_{m}\rho_{m} - \rho_{m}\hat{a}_{m}^{\dag}\hat{a}_{m}\hat{a}_{m}^{\dag}\hat{a}_{m}
\right),
\end{equation}
with $\hat{a}_{m}$ being the annihilation operator for atoms in the middle well, $m=(n+1)/2$ and $\Gamma$ is the dephasing rate.

For the numerical investigation of many-body interacting quantum systems that are too large
for master equation methods, the preferred option is the positive-P representation~\cite{PDD},
which is in principal exact, but becomes unstable
after very short times for this and many other nonlinear systems.
We therefore chose to employ stochastic integration in the truncated
Wigner representation~\cite{Robert}, which enables us to capture the majority of the quantum features of
the system and also has the huge operational advantage of remaining stable over relatively
long integration times~\cite{Sarah}. One real advantage of this method is that we can
easily add more wells with only a linear increase in computational complexity, which allows us to easily model eleven wells with a total of $2000$ atoms.  We are also able to model different initial quantum states~\cite{BECstat,BECstatPRA,BECFock}, using the methods described in Olsen and Bradley~\cite{states}.

Following the standard methods~\cite{SMCrispin,QNoise,DFW}, a mapping of our Hamiltonians onto partial differential equations for the Wigner function leads to generalised Fokker-Planck equations with third-order derivatives. Although methods have been developed to map these onto stochastic difference equations~\cite{nossoEPL,P++}, integration of the resulting equations is highly divergent. For this reason, we truncate the generalised Fokker-Planck equation at second order, which leads to stochastic differential equations which have noise terms from the dephasing term. The quantum statistics of the initial states are included in these equations via sampling the initial conditions for each stochastic trajectory from the appropriate distribution. This procedure has been shown to give accurate results for many systems~\cite{BobWig}, but is known to fail to predict revivals in the quadratures of the anharmonic oscillator, as well as giving incorrect results for two-time correlation functions~\cite{turco}. For the parameters of our investigation here, we are fully confident that the truncated Wigner procedure will give accurate results.

As an illustrative example, the truncated Wigner equations of motion for our three-well system in the Stratonovich form~\cite{SMCrispin} are
\begin{eqnarray}
\frac{d\alpha_{1}}{dt} &=& -2\chi |\alpha_{1}|^{2}\alpha_{1}+iJ\alpha_{2}, \nonumber \\
\frac{d\alpha_{2}}{dt} &=& -2\chi |\alpha_{2}|^{2}\alpha_{2}+iJ(\alpha_{1}+\alpha_{3})+i\sqrt{\Gamma}\,\alpha_{2}\eta, \nonumber \\
\frac{d\alpha_{3}}{dt} &=& -2\chi |\alpha_{3}|^{3}\alpha_{3}+iJ\alpha_{2},
\label{eq:Strat3}
\end{eqnarray}
where $\eta$ is a Gaussian noise term with the moments $\overline{\eta(t)}=0$ and $\overline{\eta(t)\eta(t')}=\delta(t-t')$. The equations for greater numbers of wells are the obvious extensions of these.
The advantage of Stratonovich calculus is that we can use higher order algorithms than is possible with the It\^o form of the equations. 
In this work, we have used the RK4
algorithm (fourth-order Runge-Kutta) with the XMDS equation solver~\cite{xmds,xmds1}. 
The $\alpha_{j}$ are the Wigner variables corresponding to the $\hat{a}_{j}$, in the sense that the average of products of the Wigner variables become equal to the expectation values of symmetrically ordered operator products in the limit of a large number of stochastic trajectories.

\section{Quantum Correlations}
\label{sec:correlations}

As well as the populations in each well, in the present system we are interested in the possibility of
entanglement between the atomic modes at each site. This is not the same as entanglement between individual atoms.
While this seems obvious because the atoms are indistinguishable bosons and only the spatially
separated modes can be distinguished, we state it here to avoid any possible confusion. In
particular, we wish to investigate entanglement between the two wells each side of the middle well. 

We will use two different correlation functions for this. From the Hillery-Zubairy inequality~\cite{HZ} we define a correlation function
\begin{equation}
\xi_{ij} = \langle \hat{a}_{i}^{\dag}\hat{a}_{j}\rangle \langle \hat{a}_{j}^{\dag}\hat{a}_{i}\rangle - \langle \hat{a}_{i}^{\dag}\hat{a}_{i}\hat{a}_{j}^{\dag}\hat{a}_{j}\rangle,
\label{eq:xi}
\end{equation}
with a positive result for this function indicating mode entanglement. However, for the systems under investigation we found no positive values of $\xi_{ij}$.
However, Dalton \etal~\cite{waffle} have since shown that, for states obeying super-selection rules, the first term on the right-hand side of Eq.~\ref{eq:xi} being greater than zero can indicate entanglement, which is a less stringent requirement. We will also therefore define the two-mode coherence function, a positive value of which may indicate entanglement,
\begin{equation}
\sigma_{ij} = \langle \hat{a}_{i}^{\dag}\hat{a}_{j}\rangle \langle \hat{a}_{j}^{\dag}\hat{a}_{i}\rangle.
\label{eq:sigmaij}
\end{equation}
We will show below that care needs to be taken with the use of this function.

Another quantity of interest is the atomic current into the middle well, defined as 
\begin{equation}
I_{m} = -i\langle \hat{a}_{m-1}^{\dag}\hat{a}_{m}-\hat{a}_{m}^{\dag}\hat{a}_{m-1}+\hat{a}_{m+1}^{\dag}\hat{a}_{m}-\hat{a}_{m}^{\dag}\hat{a}_{m+1}  \rangle,
\label{eq:current}
\end{equation}
which is simple to calculate in the truncated WIgner representation.

The last quantity we define here is a pseudo-entropy of the system~\cite{BHrelax,AnglinVardi}, calculated from the reduced single-particle density matrix. The latter is defined for three wells as
\begin{equation}
{\cal R}_{3} = \frac{1}{\sum_{i=1}^{3}\langle \hat{a}_{i}^{\dag}\hat{a}_{i}\rangle}
\begin{bmatrix}
 \langle\hat{a}_{1}^{\dag}\hat{a}_{1}\rangle & \langle\hat{a}_{1}^{\dag}\hat{a}_{2}\rangle & \langle\hat{a}_{1}^{\dag}\hat{a}_{3}\rangle \\
\langle\hat{a}_{2}^{\dag}\hat{a}_{1}\rangle & \langle\hat{a}_{2}^{\dag}\hat{a}_{2}\rangle & \langle\hat{a}_{2}^{\dag}\hat{a}_{3}\rangle \\
\langle\hat{a}_{3}^{\dag}\hat{a}_{1}\rangle & \langle\hat{a}_{3}^{\dag}\hat{a}_{2}\rangle & \langle\hat{a}_{3}^{\dag}\hat{a}_{3}\rangle \\
 \end{bmatrix},
\label{eq:Rmat}
\end{equation}
with the extension to higher numbers of wells, ${\cal R}_{n}$, being obvious. The advantage of this reduced density matrix is that all the quantities needed are simple to calculate in the truncated Wigner representation, with the computational expense being largely independent of the number of atoms, and linearly dependent on the number of wells.
The pseudo-entropy is then defined in the standard von Neumann manner 
as
\begin{equation}
\zeta_{n} = -\tr\left({\cal R}_{n}\log{\cal R}_{n}\right),
\label{eq:vNS}
\end{equation}
where $\log x$ is the natural logarithm of $x$.
Analytical values for $\zeta_{n}$ can be calculated in some limiting cases, such as systems of Fock or coherent states, and also for equal populations with no intermode coherences. These limiting cases are useful for the calculation of maximum values to which the system should relax if all coherences disappear and the populations are equal, where $\zeta_{n}=\log n$,  and show how close to the expected equilibrium state the system will approach. As a final note, we mention that all the quantities needed for the correlations above can in principle be measured, either by density (number) measurements or via atomic homodyning~\cite{andyhomo}. 

\section{Results}
\label{sec:results}

For all our simulations, we set $J=1$ and $\chi = 0.01$, with the middle well initially empty and all others containing an average of $200$ atoms.
As well as considering a dephasing parameter of $\Gamma=0$ or $1.5$, we have considered initial Fock states and coherent states in the initially occupied wells. A Fock state of fixed number is the natural state for atoms in an isolated well, while coherent states are more natural in the superfluid regime. The new developments in the manipulation of potentials make both these initial states possible. Fock states could be obtained in isolated wells which were then moved into proximity so that tunnelling could begin. Coherent states would be more natural if an optical lattice were imposed on an already condensed Bose gas, in which case the central well could be depleted using the electron beam technique~\cite{NDC}.
We have limited the number of wells investigated to a maximum of $11$ because the observables in the middle well do not depend noticeably on the total number once the chains approach this length. The
numerical solutions were averaged over at least $10^{6}$ stochastic realisations of the truncated Wigner equations and sampling errors are less than the line widths.

\subsection{Three wells}
\label{subsec:trimer}

\begin{figure}[tbhp]
\includegraphics[width=0.75\columnwidth]{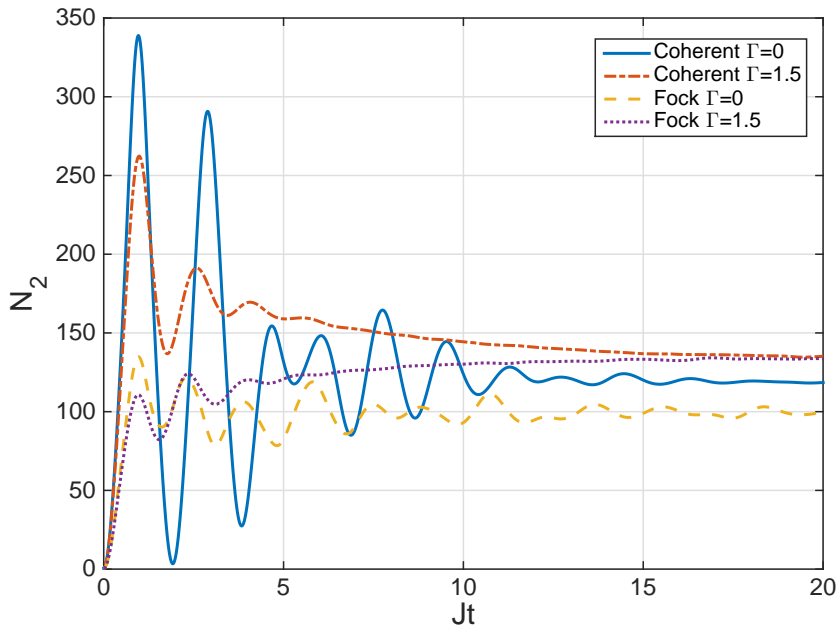}
\caption{(colour online) The numerical solutions for the population in the middle well as a function of time for both coherent
and Fock initial states, with different dephasing rates, for the trimer. The parameter values are $J=1$, $\chi=10^{-2}$, $N_{1}(0)=N_{3}(0)=200$, and $N_{2}(0)=0$. $Jt$ is a dimensionless time and all quantities plotted in this and subsequent graphs are dimensionless.}
\label{fig:N2triwell}
\end{figure}

\begin{figure}[tbhp]
\includegraphics[width=0.75\columnwidth]{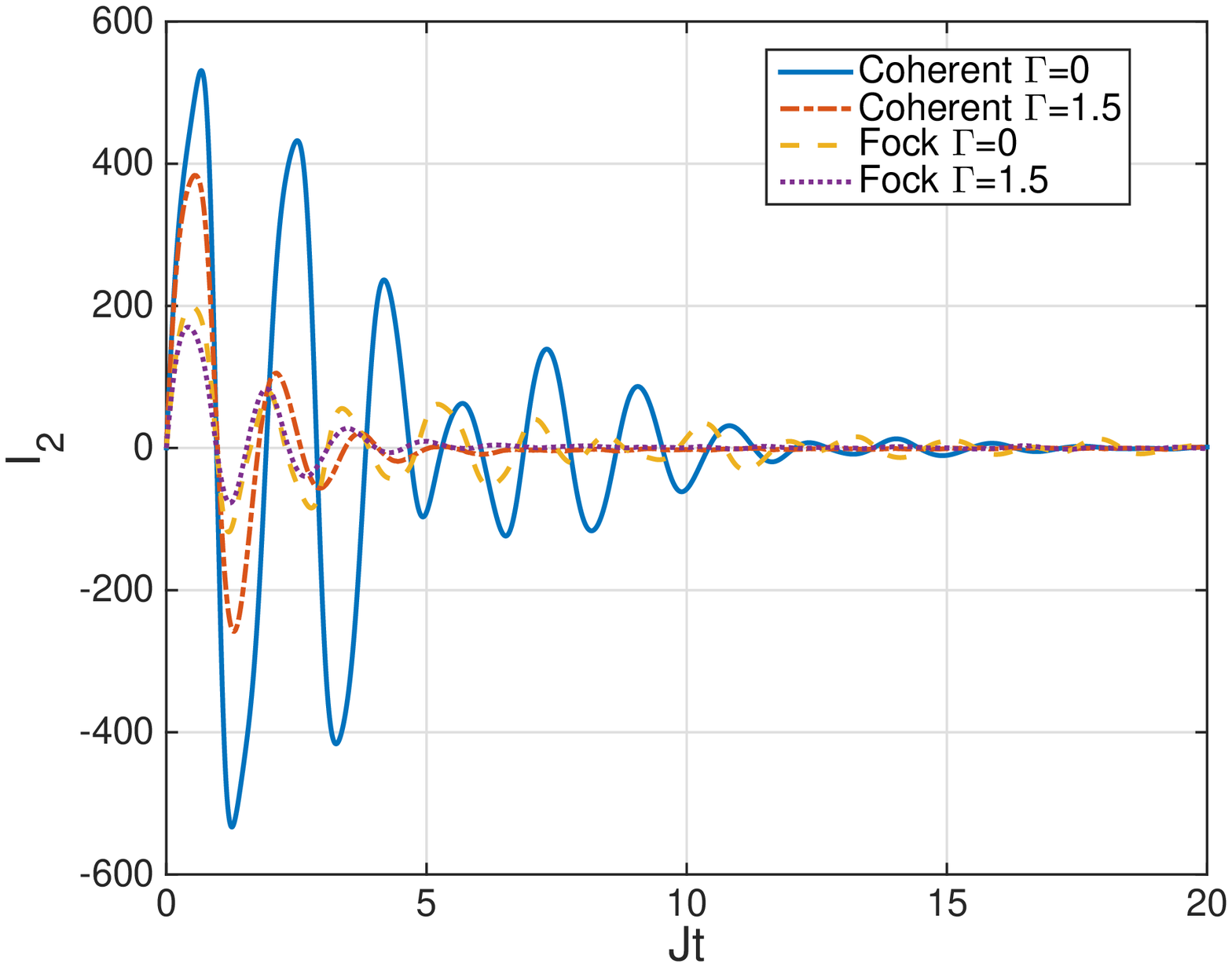}
\caption{(colour online) The numerical solutions for the currents into the middle well of the trimer as a function of time for both coherent
and Fock initial states, with different dephasing rates. The parameter values are as in Fig.~\ref{fig:N2triwell}.}
\label{fig:I2}
\end{figure}

This is analogous to the system investigated by Olsen and Corney~\cite{ourNDC} with respect to negative differential conductivity, albeit with larger numbers of atoms in the two end wells. Firstly we investigated the effects of dephasing and initial quantum states on the evolution of the population in the middle well, with the results shown in Fig.~\ref{fig:N2triwell}. We immediately see that the evolution without dephasing is markedly dependent on the initial quantum states. This is to be expected since the tunneling operator is a function of coherences between the modes. We also see that the populations approach different stationary values. Once the dephasing is included, tunnelling for both states is slowed, with fewer oscillations in the populations, and steady-state values with the atoms distributed evenly between the three wells are reached. The dynamics of the current into the middle well are shown in Fig.~\ref{fig:I2}, from which we see that initial coherent states cause larger amplitude oscillations than initial Fock states, and that tunnelling is damped for both initial states by the inclusion of dephasing, with the steady-state tunnelling going to zero for all initial conditions, whether dephasing is included or not.

\begin{figure}[tbhp]
\includegraphics[width=0.75\columnwidth]{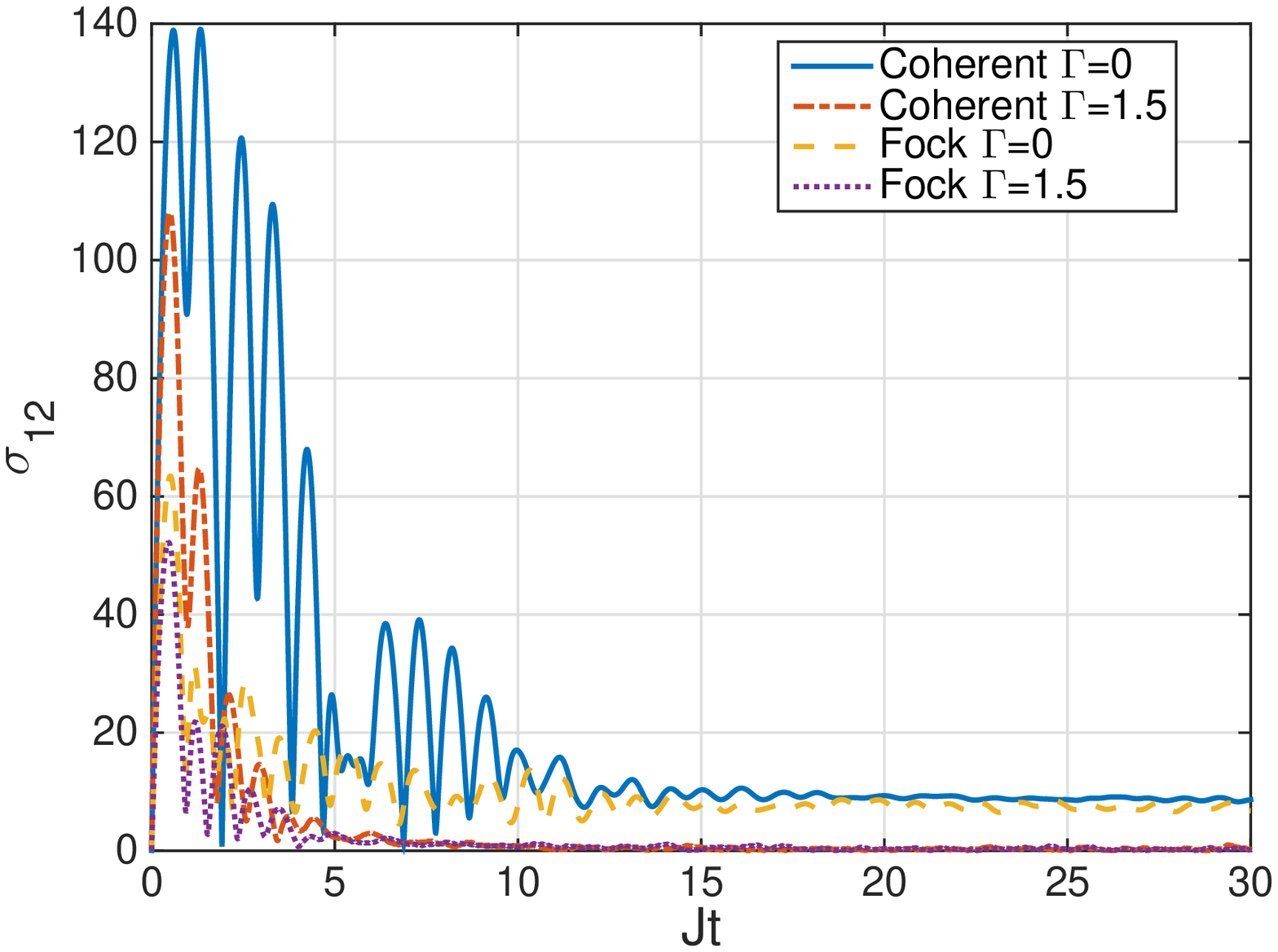}
\caption{(colour online) The numerical solutions for $\sigma_{12}$ in the three-well system, as a function of time for both coherent and Fock initial states, for the same parameters as Fig.~\ref{fig:N2triwell}.}
\label{fig:sigma12}
\end{figure}

\begin{figure}[tbhp]
\includegraphics[width=0.75\columnwidth]{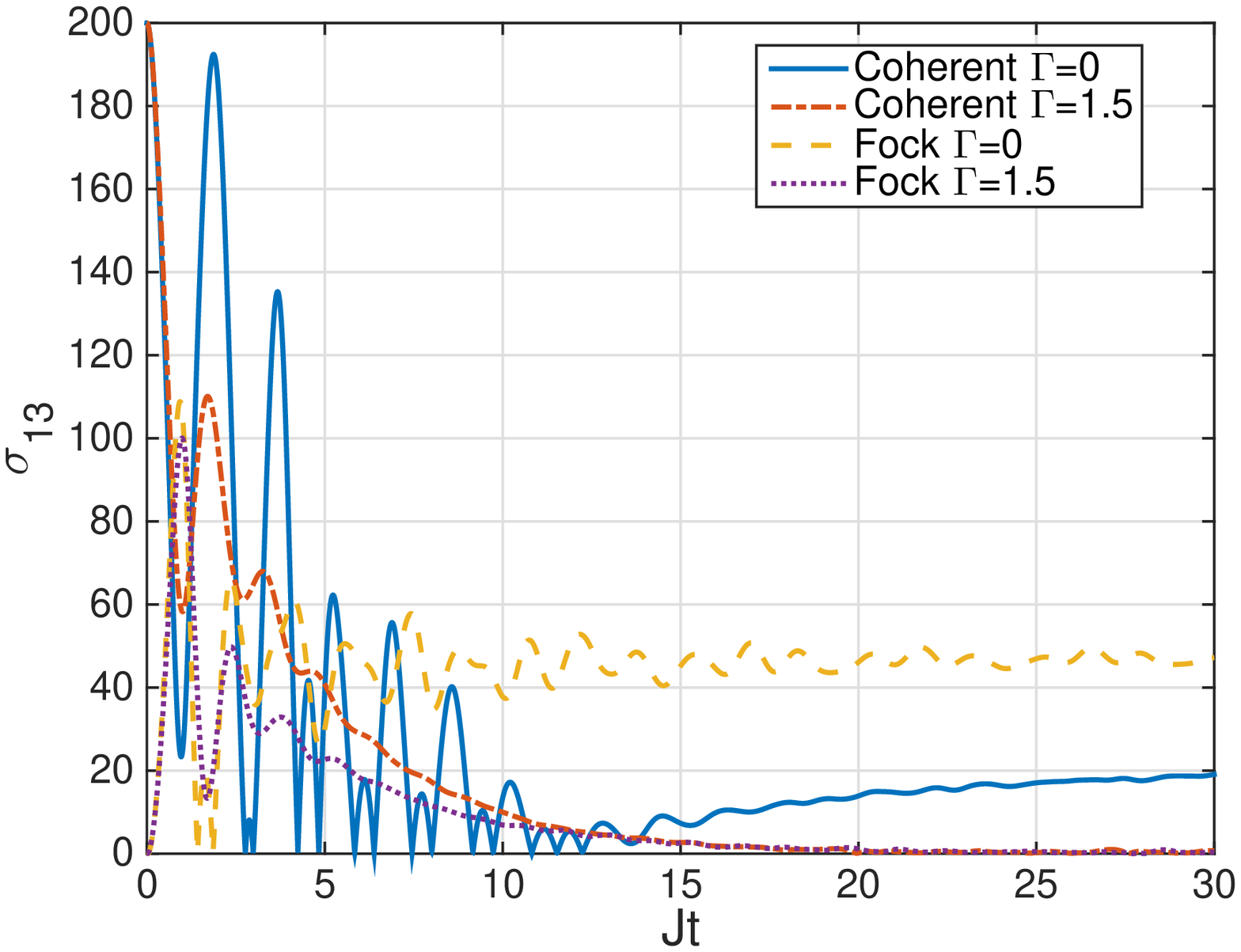}
\caption{(colour online) The numerical solutions for $\sigma_{13}$ in the three-well system, as a function of time for both coherent and Fock initial states, for the same parameters as Fig.~\ref{fig:N2triwell}.}
\label{fig:sigma13}
\end{figure}

We also investigated the $\xi_{ij}$ correlations of Eq.~\ref{eq:xi}, but found no positive values for the parameters used. However, this correlation is sufficient but not necessary to denote entanglement between modes $i$ and $j$, and as shown by Dalton \etal~\cite{waffle}, a positive value of the $\sigma_{ij}$ correlation of Eq.~\ref{eq:sigmaij} can also indicate entanglement between the relevant modes. As seen in Fig.~\ref{fig:sigma12} and Fig.~\ref{fig:sigma13}, we found positive steady-state values for both $\sigma_{12}$ and $\sigma_{13}$ without dephasing, for both Fock and coherent initial states. $\sigma_{12}$ detects entanglement between the first well and the middle well in this case, while $\sigma_{13}$ detects entanglement between the two wells each side of the middle, initially depleted well. We note here that, due to the symmetry of the system, $\sigma_{12}=\sigma_{23}$. 
In the cases below, with a greater number of wells, we will always investigate the coherence between the middle well and the one to its left, as well as between the two adjacent to the middle. 
The inclusion of $\Gamma=1.5$ in our equations destroyed the steady-state coherences and hence the entanglement in both cases, which is to be expected since this dephasing destroys the off-diagonal elements of the system density matrix without changing the number distribution~\cite{QNoise}. The results shown here for $\sigma_{13}$ show that this correlation can be misleading. Similarly to the generalised purity~\cite{OCChianca}, a naive examination of the results indicates that wells $1$ and $3$ with initial coherent states are entangled at $t=0$, despite never having interacted. This is contrary to the spirit of entanglement as defined by Schrödinger~\cite{Ernie}, who said “When two systems,
of which we know the states by their respective representatives,
enter into temporary physical interaction due to known forces between
them, and when after a time of mutual influence the systems
separate again, then they can no longer be described in the same
way as before, viz. by endowing each of them with a representative
of its own. I would not call that one but rather the characteristic
trait of quantum mechanics, the one that enforces its entire departure
from classical lines of thought. By the interaction the two representatives
[the quantum states] have become entangled”. We note that, for initial Fock states, this measure begins at zero, as it must to be a true entanglement measure. In view of the demonstrated unreliability of this measure, we will examine the pseudo-entropy of Eq.~\ref{eq:vNS} as an entanglement measure, in Section~\ref{sec:pseudo}.

\subsection{Five wells and more}
\label{subsec:cinco}

\begin{figure}[tbhp]
\includegraphics[width=0.75\columnwidth]{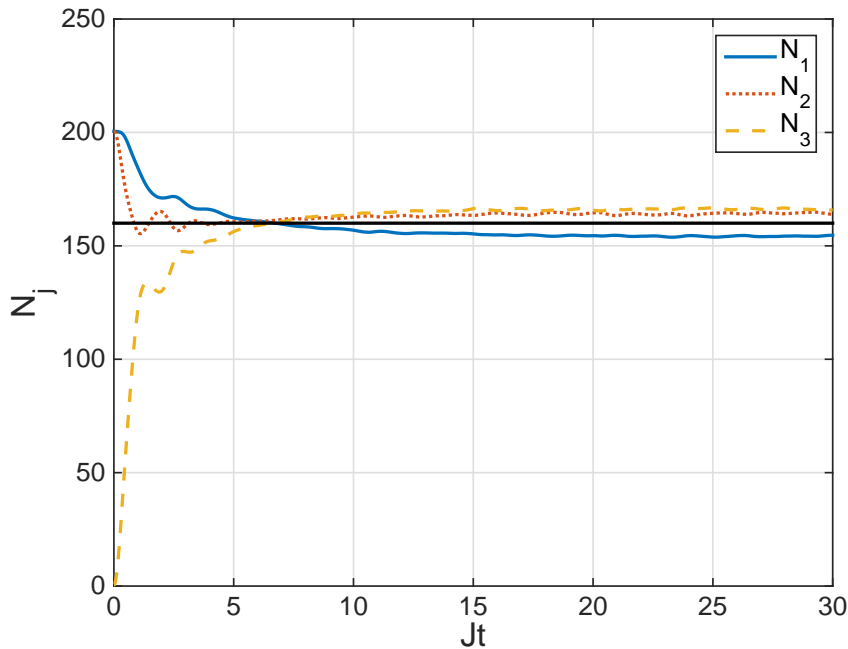}
\caption{(colour online) The populations $N_{1}$, $N_{2}$ and $N_{3}$ for initial Fock states in the five-well system with $\Gamma=1.5$. Due to the symmetry of the system, $N_{5}=N_{1}$ and $N_{4}=N_{2}$. The horizontal line is a guide to the eye at the values expected if all wells finished with equal populations. The initial conditions were $N_{1}(0)=N_{2}(0)=N_{4}(0)=N_{5}(0)$ and $N_{3}(0)=0$, with $\chi=10^{-2}$.}
\label{fig:Fock5}
\end{figure}

\begin{figure}[tbhp]
\includegraphics[width=0.75\columnwidth]{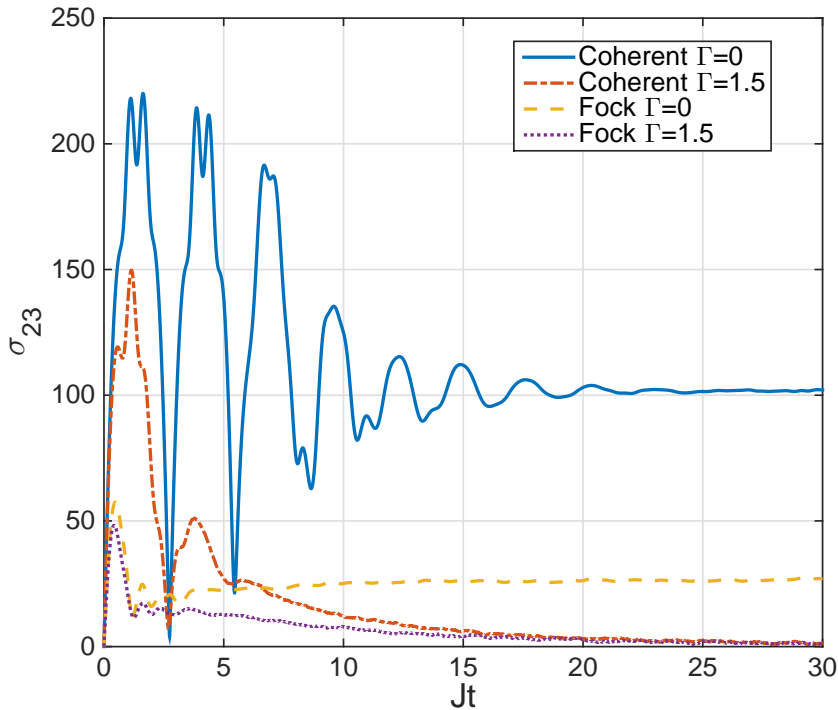}
\caption{(colour online) The coherence function, $\sigma_{23}$, between the middle well and the one to its left for the five-well chain. Initial conditions and parameters are as for Fig.~\ref{fig:Fock5}.}
\label{fig:Sigma23}
\end{figure}

Once we examine chains longer than three wells, we find that the populations do not equilibrate over the whole chain for longer times except for $\Gamma=0$ and initial Fock states, with the end wells having less steady-state populations than those in the interior of the chain. This is shown for initial Fock states in Fig.~\ref{fig:Fock5}, where the steady-state population of the first well is less than that of wells $2$ and $3$. This was also found to be a feature of longer chains, with the end wells obviously existing in a different environment to those in the interior of the chain. We expect that this would not occur for a ring of Bose-Hubbard wells, where all would exist in the same dynamical environment. Because the dynamics, particularly those of the middle well, did not change appreciably for the addition of more than two wells each side of the centre, we will not present results for higher numbers of wells graphically. The pseudo-entropy, which did change, is addressed in Section~\ref{sec:pseudo}.

For $n\geq 5$, we also did not find any evidence of entanglement using the Hillery-Zubairy criteria. On the other hand, as shown in Fig.~\ref{fig:Sigma23}, the coherence correlation $\sigma_{23}$ increases from zero at the beginning of the evolution to settle to a steady-state value. Without dephasing, the steady-state values are greater than zero. With dephasing, $\sigma_{23}$ decays to zero after some initial oscillatory behaviour. Once again, this shows that dephasing destroys coherence and entanglement. We note again that the results for $\sigma_{24}$ and initial coherent states also emphasised the care that must be taken with this measure if it is to be used as an indicator of entanglement, since positive values were again found for $t=0$.

\section{Pseudo-entropy}
\label{sec:pseudo}

For purposes of easy comparison, we present the results for the pseudo-entropies jointly, in this section, for chains ranging from $3$ to $11$ wells. We begin with the pseudo-entropy of the whole chain, as defined by Eq.~\ref{eq:vNS}, calculating this for five different chain lengths, and for initial Fock and coherent states, with $\Gamma=0$ and $\Gamma=1.5$.

\begin{figure}[tbhp]
\includegraphics[width=0.75\columnwidth]{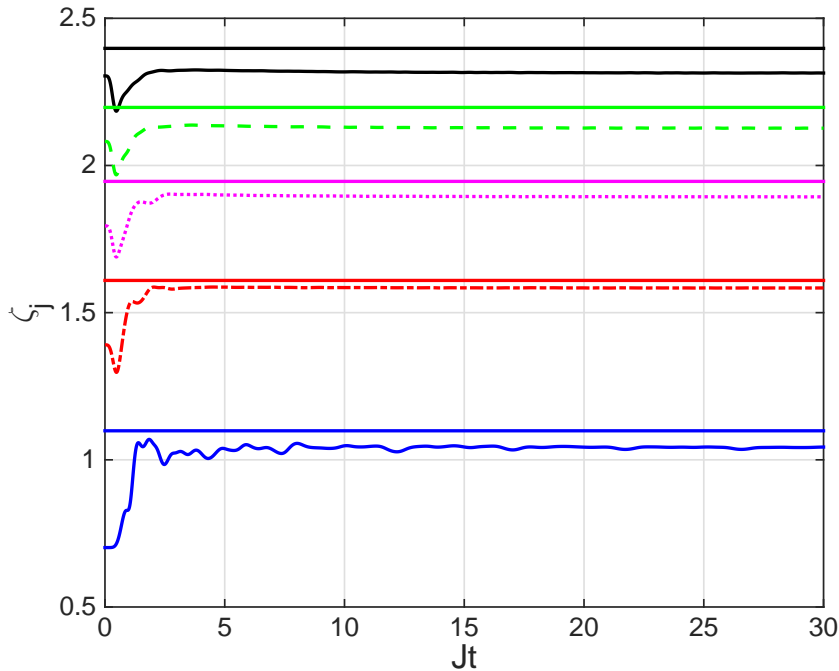}
\caption{(colour online) The pseudo-entropies, $\zeta_{n}$, for initial Fock states with $\Gamma=0$, $\chi=10^{-2}$, and different chain lengths. $\zeta_{3}$ is the solid blue line, $\zeta_{5}$ is the dot-dashed red line, $\zeta_{7}$ is the dotted magenta line, $\zeta_{9}$ is the dashed green line, and $\zeta_{11}$ is the solid black line. The solid horizontal lines are the maximum entropy expected for a chain in full equilibrium, and are a guide to the eye.}
\label{fig:ZetaF0}
\end{figure}

\begin{figure}[tbhp]
\includegraphics[width=0.75\columnwidth]{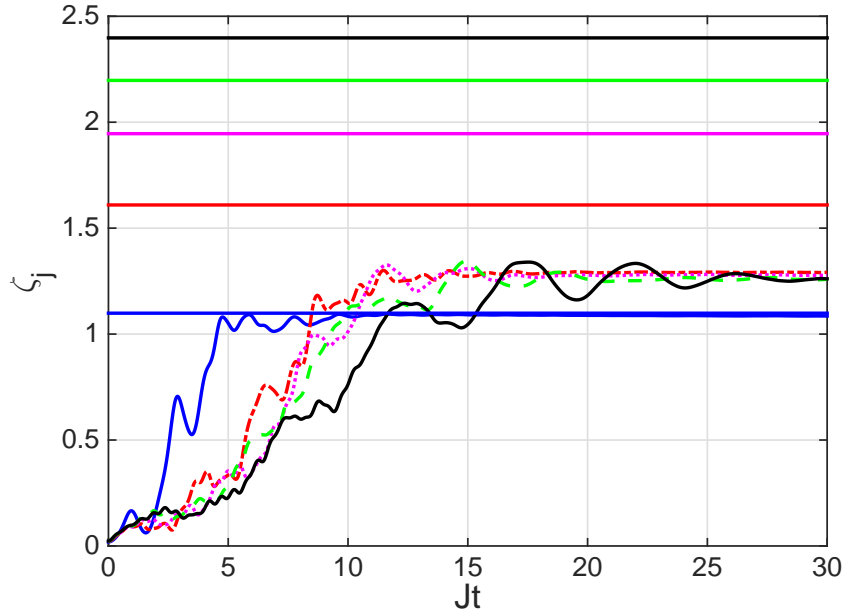}
\caption{(colour online) The pseudo-entropies, $\zeta_{n}$, for initial coherent states with $\Gamma=0$ and different chain lengths. $\zeta_{3}$ is the solid blue line, $\zeta_{5}$ is the dot-dashed red line, $\zeta_{7}$ is the dotted magenta line, $\zeta_{9}$ is the dashed green line, and $\zeta_{11}$ is the solid black line. The solid horizontal lines are the maximum entropy expected for a chain in full equilibrium, and are a guide to the eye.}
\label{fig:ZetaC0}
\end{figure}

The first thing we observe from the figures is that, without phase diffusion, none of the configurations reach the expected value, $\log n$, of the entropy for a totally equilibrated chain, where the density matrix would be diagonal. As shown in Fig.~\ref{fig:ZetaF0}, for initial Fock states the results are all below $\log n$.  With initial coherent states, as shown in Fig.~\ref{fig:ZetaC0}, this behaviour is even more obvious.  We see that only the three-well chain approaches full equilibrium without phase diffusion, whereas the longer chains all converge on a similar value, well below their respective maxima. This demonstrates that making the chain longer than five wells does not change the equilibrium state without phase diffusion.

\begin{figure}[tbhp]
\includegraphics[width=0.75\columnwidth]{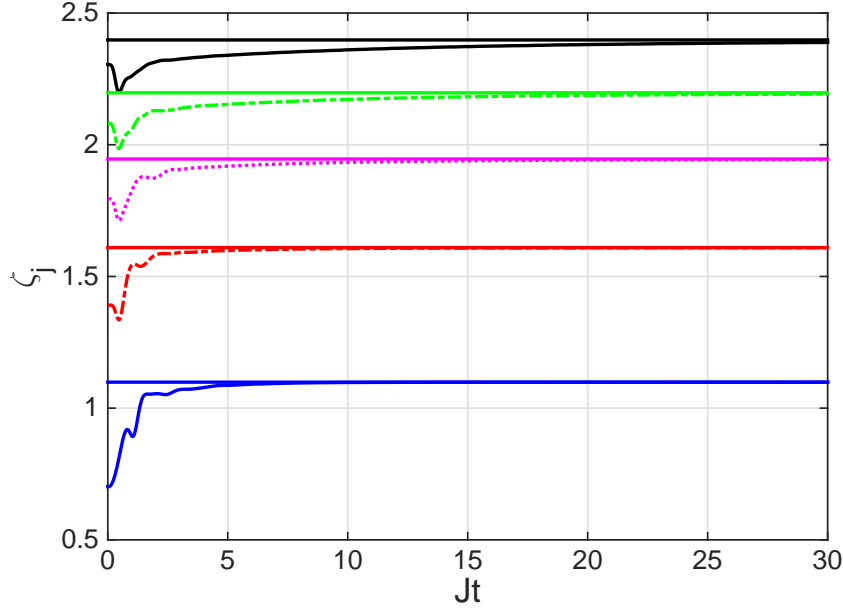}
\caption{(colour online) The pseudo-entropies, $\zeta_{n}$, for initial Fock states with $\Gamma=1.5$ and different chain lengths. $\zeta_{3}$ is the solid blue line, $\zeta_{5}$ is the dot-dashed red line, $\zeta_{7}$ is the dotted magenta line, $\zeta_{9}$ is the dashed green line, and $\zeta_{11}$ is the solid black line. The solid horizontal lines are the maximum entropy expected for a chain in full equilibrium, and are a guide to the eye. }
\label{fig:ZetaF15}
\end{figure}

\begin{figure}[tbhp]
\includegraphics[width=0.75\columnwidth]{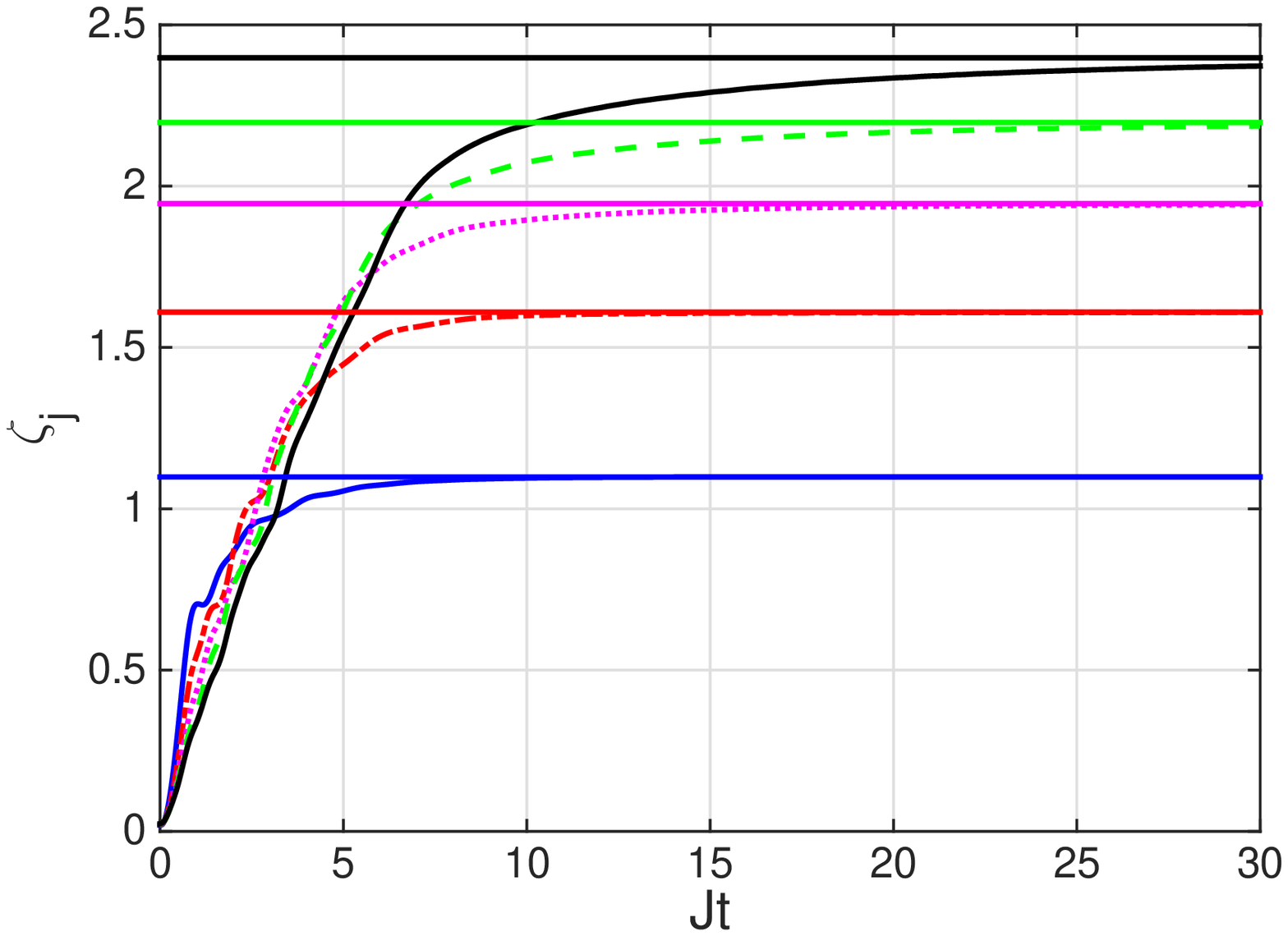}
\caption{(colour online) The pseudo-entropies, $\zeta_{n}$, for initial coherent states with $\Gamma=1.5$ and different chain lengths. $\zeta_{3}$ is the solid blue line, $\zeta_{5}$ is the dot-dashed red line, $\zeta_{7}$ is the dotted magenta line, $\zeta_{9}$ is the dashed green line, and $\zeta_{11}$ is the solid black line. The solid horizontal lines are the maximum entropy expected for a chain in full equilibrium, and are a guide to the eye.}
\label{fig:ZetaC15}
\end{figure}

However, when we include a finite $\Gamma$, all the chains considered relax to close to their equilibrium values, as shown in Fig.~\ref{fig:ZetaF15} and Fig.~\ref{fig:ZetaC15}, although longer chains take more time to reach this state. These maximum pseudo-entropy results seem to indicate that there is no entanglement present between the wells once phase diffusion is included, which would be inconsistent with the results found above for the $\sigma_{ij}$. However, we have already found that this measure can be unreliable. 

The behaviour of $\zeta_{n}$ can be understood more fully when we consider the reduced single-particle density matrices at the final evolution time. In full equilibrium, all off-diagonal elements will have decayed to zero, and the diagonal elements will be equal. Considering the three-well chain at $Jt=30$, when we look at $R_{3}$ for initial coherent states with $\Gamma=0$, we find
\begin{equation}
{\cal R}_{3} = 
\begin{bmatrix}
0.3538 & -0.0216+0.0008i & -0.0475+0.003i \\
-0.0216-0.0008i & 0.2924 & -0.0218-0.0005i \\
 -0.0475-0.0003i & -0.0218+0.0005i  & 0.3528 \\
 \end{bmatrix},
\label{eq:RmatC0}
\end{equation}
whereas with $\Gamma=1.5$, we find
\begin{equation}
{\cal R}_{3} = 
\begin{bmatrix}
0.3321 & -0.0001+0.0008i  & 0.0003+0.0016i \\
-0.0001-0.0008i & 0.3341  & 0.0003+0.0001i  \\
 0.0003-0.0016i & 0.0003-0.0001i  & 0.3338 \\
 \end{bmatrix}.
\label{eq:RmatC15}
\end{equation}
We can see directly from this how the dephasing has acted to destroy the off-diagonal coherences. In the case of initial Fock states with $\Gamma=0$, we find
\begin{equation}
{\cal R}_{3} = 
\begin{bmatrix}
0.3739 & -0.0168+0.0010i & -0.1176-0.0004i \\
-0.0168-0.0010i & 0.2530 & -0.0178-0.008i \\
-0.1176+0.0004i & -0.0178+0.0008i & 0.3731 \\
 \end{bmatrix},
\label{eq:RmatF0}
\end{equation}
which shows coherences which have built up during the evolution of the system, since they are always zero for two independent Fock states, which is the initial condition. For Fock states and $\Gamma=1.5$, we find
\begin{equation}
{\cal R}_{3} = 
\begin{bmatrix}
0.3323 & 0.0001+0.0002i & -0.0008-0.0002i \\
0.0001-0.0002i & 0.3343 & 0.0010+0.0006i \\
-0.0008+0.0002i & 0.0010-0.0006i & 0.3334 \\
\end{bmatrix},
\label{eq:RmatF15}
\end{equation}
which again shows that, in the competition between build up of coherence via tunneling and destruction of coherence via dephasing, the dephasing is having more of an effect and the chain has relaxed to something close to its equilibrium state.

\begin{figure}[tbhp]
\includegraphics[width=0.75\columnwidth]{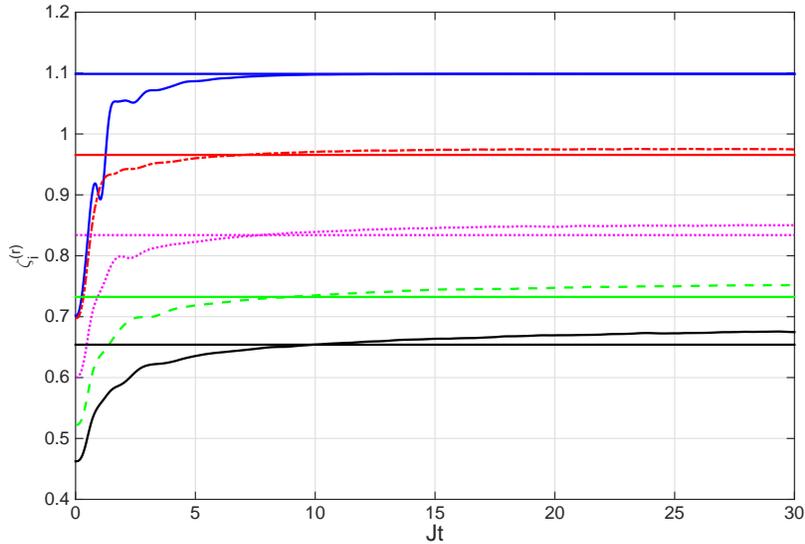}
\caption{(colour online) The reduced pseudo-entropies, $\zeta_{n}^{(r)}$, for initial Fock states with $\Gamma=1.5$ and different chain lengths. $\zeta_{3}^{(r)}$ is the solid blue line, $\zeta_{5}^{(r)}$ is the dot-dashed red line, $\zeta_{7}^{(r)}$ is the dotted magenta line, $\zeta_{9}^{(r)}$ is the dashed green line, and $\zeta_{11}^{(r)}$ is the solid black line. The solid horizontal lines are the maximum entropy expected for a chain in full equilibrium, and are a guide to the eye.}
\label{fig:midzetaF15}
\end{figure}

\begin{figure}[tbhp]
\includegraphics[width=0.75\columnwidth]{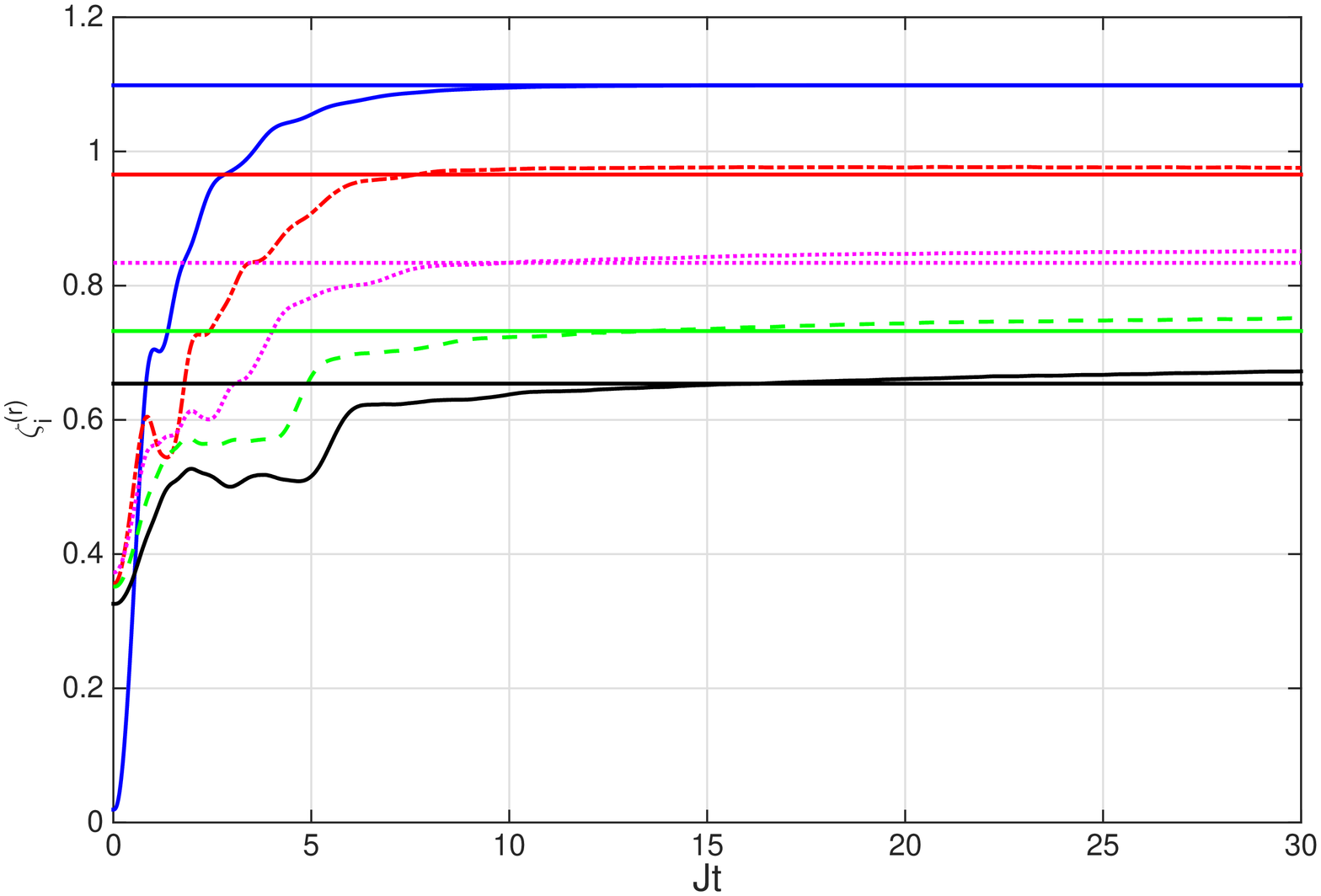}
\caption{(colour online) The reduced pseudo-entropies, $\zeta_{n}^{(r)}$, for initial coherent states with $\Gamma=1.5$ and different chain lengths. $\zeta_{3}^{(r)}$ is the solid blue line, $\zeta_{5}^{(r)}$ is the dot-dashed red line, $\zeta_{7}^{(r)}$ is the dotted magenta line, $\zeta_{9}^{(r)}$ is the dashed green line, and $\zeta_{11}^{(r)}$ is the solid black line. The solid horizontal lines are the maximum entropy expected for a chain in full equilibrium, and are a guide to the eye.}
\label{fig:midzetaC15}
\end{figure}

It is also useful here to define a reduced density matrix for the middle three wells of an $n$-well chain,
\begin{equation}
\tilde{{\cal R}}_{3}^{n} = \frac{1}{\sum_{i=1}^{n}\langle \hat{a}_{i}^{\dag}\hat{a}_{i}\rangle}
\begin{bmatrix}
 \langle\hat{a}_{j-1}^{\dag}\hat{a}_{j-1}\rangle & \langle\hat{a}_{j-1}^{\dag}\hat{a}_{j}\rangle & \langle\hat{a}_{j-1}^{\dag}\hat{a}_{j+1}\rangle \\
\langle\hat{a}_{j}^{\dag}\hat{a}_{j-1}\rangle & \langle\hat{a}_{j}^{\dag}\hat{a}_{j}\rangle & \langle\hat{a}_{j}^{\dag}\hat{a}_{j+1}\rangle \\
\langle\hat{a}_{j+1}^{\dag}\hat{a}_{j-1}\rangle & \langle\hat{a}_{j+1}^{\dag}\hat{a}_{j}\rangle & \langle\hat{a}_{j+1}^{\dag}\hat{a}_{j+1}\rangle \\
 \end{bmatrix},
\label{eq:R3mat}
\end{equation}
where $j=(n-1)/2$, and hence a reduced three-well pseudo-entropy,
\begin{equation}
\zeta_{n}^{(r)} = -\tr\left(\tilde{{\cal R}}_{3}^{n}\log{\tilde{\cal R}}_{3}^{n}\right).
\label{eq:vNSreduce}
\end{equation}
This value gives the reduced pseudo-entropy of the middle three-well subsystem, and it is also simple to calculate the maximum equilibrium values, when populations are equilibrated along the chain and all coherences have disappeared. In Fig.~\ref{fig:midzetaF15} we show the reduced subsystem pseudo-entropies for the different chain lengths, with initial Fock states and $\Gamma=1.5$. What we see is that, once we have more than one well either side of the middle, the subsystem entropy becomes higher than we would expect from an equilibrium situation. This was not the case for the total entropy. 

However, remembering that with more than three wells, the end wells have lower occupations than the middle ones at the end of our evolution times, it is instructive to look at the actual occupations of the three middle wells and calculate the reduced pseudo-entropies for the situation with these actual populations and all off-diagonal elements being zero. As an illustrative example, we have done this for $n=7$. Beginning with initial Fock states with $\chi=10^{-2}$, we find a value of $\zeta_{7}^{(r)} = 0.83$ if the well populations are equal. This is shown as the horizontal dotted line in Fig.~\ref{fig:midzetaF15}. However, the actual long-time populations found by stochastic integration are not all equal. We found $N_{3} = N_{5} = 178$ and $N_{4} =189$. If the populations were all equal, we would expect $N_{j}=171\,\forall j$. With no off-diagonal elements in the density matrix, these populations give a value of $\zeta_{7}^{(r)}=0.86$, which is close to that shown in Fig.~\ref{fig:midzetaF15}. A similar check for initial coherent states gives $N_{3} = N_{5} = 179$ and $N_{4} =189$. The reduced pseudo-entropy calculated with these numbers also closely matched that actually found. We therefore conclude that the increased values of the pseudo-entropy for the three middle wells as compared to that of a full equilibrium situation are due to the increased populations in the central wells. As soon as we have more than three wells, this begins to happen, and is a finite size effect due to the different dynamics of the end wells. It is not evidence of entropy of entanglement. 

\section{Conclusions}
\label{sec:conclusions}

We have investigated the influence of phase diffusion on different size Bose-Hubbard chains with
the middle well initially empty. We have considered the influence on
population equilibration, on mode entanglement and on the current into the middle well. We have found that, except in the simplest three-well case, the populations do not equilibrate, with the wells close to the middle holding higher numbers of atoms in the steady-state than those nearer the ends. We have found no conclusive evidence of mode entanglement, with the Hillery-Zubairy criterion obtaining no positive values. The simpler criterion based on the coherence function has been shown to give unreliable results. The reduced entanglement of the middle three wells being higher than the equilibrium values is explained by the population imbalance in the middle wells, and cannot be read as entropy of entanglement.

We have shown that the presence of phase diffusion affects the atomic currents into the middle well, although this does not change qualitatively once we have more than five wells in the chain. This effect is the same as that previously discovered in investigations of negative differential conductivity. The phase diffusion, as expected, acts to destroy coherences between the modes.

We have found that the two parameters which affect the dynamics most markedly are the initial quantum states and the rate of phase diffusion. The initial quantum states can be controlled experimentally by the preparation of an experiment, while the phase diffusion rate could in principle be changed by the depth of the middle well. The most surprising discovery is perhaps how small a chain can be before finite size effects from the ends stop having a noticeable effect on the dynamics of the central three wells. What we cannot claim here is that this conclusion is general, as it may depend in parameters such as the collisional nonlinearity and the initial numbers of atoms.

\section*{Acknowledgments}
This research was supported by the Australian Research Council under the Future Fellowships Program (Grant ID: FT100100515), and the International Office of ENSICAEN. F.M. thanks Cinthya Chianca and Joel Corney for invaluable help with the application XMDS, and the UQ Physics Department for extremely useful discussions and hospitality.

\end{document}